\documentclass[twoside,reqno]{bjp}
\usepackage{epsfig,cite}
\usepackage{graphics}
\usepackage{amssymb,amsmath}
\usepackage{times}
\begin{document}

\sloppy \raggedbottom
\setcounter{page}{1}

%
%
%
%
\newcommand\rf[1]{(\ref{eq:#1})}
\newcommand\lab[1]{\label{eq:#1}}
\newcommand\nonu{\nonumber}
\newcommand\br{\begin{eqnarray}}
\newcommand\er{\end{eqnarray}}
\newcommand\be{\begin{equation}}
\newcommand\ee{\end{equation}}
\newcommand\eq{\!\!\!\! &=& \!\!\!\! }
\newcommand\foot[1]{\footnotemark\footnotetext{#1}}
\newcommand\lb{\lbrack}
\newcommand\rb{\rbrack}
\newcommand\llangle{\left\langle}
\newcommand\rrangle{\right\rangle}
\newcommand\blangle{\Bigl\langle}
\newcommand\brangle{\Bigr\rangle}
\newcommand\llb{\left\lbrack}
\newcommand\rrb{\right\rbrack}
\newcommand\Blb{\Bigl\lbrack}
\newcommand\Brb{\Bigr\rbrack}
\newcommand\lcurl{\left\{}
\newcommand\rcurl{\right\}}
\renewcommand\({\left(}
\renewcommand\){\right)}
\renewcommand\v{\vert}                     
\newcommand\bv{\bigm\vert}               
\newcommand\Bgv{\;\Bigg\vert}            
\newcommand\bgv{\bigg\vert}              
\newcommand\lskip{\vskip\baselineskip\vskip-\parskip\noindent}
\newcommand\mskp{\par\vskip 0.3cm \par\noindent}
\newcommand\sskp{\par\vskip 0.15cm \par\noindent}
\newcommand\bc{\begin{center}}
\newcommand\ec{\end{center}}
\newcommand\Lbf[1]{{\Large \textbf{{#1}}}}
\newcommand\lbf[1]{{\large \textbf{{#1}}}}




\newcommand\tr{\mathop{\mathrm tr}}                  
\newcommand\Tr{\mathop{\mathrm Tr}}                  
\newcommand\partder[2]{\frac{{\partial {#1}}}{{\partial {#2}}}}
\newcommand\partderd[2]{{{\partial^2 {#1}}\over{{\partial {#2}}^2}}}
\newcommand\partderh[3]{{{\partial^{#3} {#1}}\over{{\partial {#2}}^{#3}}}}
\newcommand\partderm[3]{{{\partial^2 {#1}}\over{\partial {#2} \partial{#3} }}}
\newcommand\partderM[6]{{{\partial^{#2} {#1}}\over{{\partial {#3}}^{#4}{\partial {#5}}^{#6} }}}          
\newcommand\funcder[2]{{{\delta {#1}}\over{\delta {#2}}}}
\newcommand\Bil[2]{\Bigl\langle {#1} \Bigg\vert {#2} \Bigr\rangle}  
\newcommand\bil[2]{\left\langle {#1} \bigg\vert {#2} \right\rangle} 
\newcommand\me[2]{\left\langle {#1}\right|\left. {#2} \right\rangle} 

\newcommand\sbr[2]{\left\lbrack\,{#1}\, ,\,{#2}\,\right\rbrack} 
\newcommand\Sbr[2]{\Bigl\lbrack\,{#1}\, ,\,{#2}\,\Bigr\rbrack}
\newcommand\Gbr[2]{\Bigl\lbrack\,{#1}\, ,\,{#2}\,\Bigr\} }
\newcommand\pbr[2]{\{\,{#1}\, ,\,{#2}\,\}}       
\newcommand\Pbr[2]{\Bigl\{ \,{#1}\, ,\,{#2}\,\Bigr\}}  
\newcommand\pbbr[2]{\lcurl\,{#1}\, ,\,{#2}\,\rcurl}




\renewcommand\a{\alpha}
\renewcommand\b{\beta}
\renewcommand\c{\chi}
\renewcommand\d{\delta}
\newcommand\D{\Delta}
\newcommand\eps{\epsilon}
\newcommand\vareps{\varepsilon}
\newcommand\g{\gamma}
\newcommand\G{\Gamma}
\newcommand\grad{\nabla}
\newcommand\h{\frac{1}{2}}
\renewcommand\k{\kappa}
\renewcommand\l{\lambda}
\renewcommand\L{\Lambda}
\newcommand\m{\mu}
\newcommand\n{\nu}
\newcommand\om{\omega}
\renewcommand\O{\Omega}
\newcommand\p{\phi}
\newcommand\vp{\varphi}
\renewcommand\P{\Phi}
\newcommand\pa{\partial}
\newcommand\tpa{{\tilde \partial}}
\newcommand\bpa{{\bar \partial}}
\newcommand\pr{\prime}
\newcommand\ra{\rightarrow}
\newcommand\lra{\longrightarrow}
\renewcommand\r{\rho}
\newcommand\s{\sigma}
\renewcommand\S{\Sigma}
\renewcommand\t{\tau}
\renewcommand\th{\theta}
\newcommand\bth{{\bar \theta}}
\newcommand\Th{\Theta}
\newcommand\z{\zeta}
\newcommand\ti{\tilde}
\newcommand\wti{\widetilde}
\newcommand\twomat[4]{\left(\begin{array}{cc}  
{#1} & {#2} \\ {#3} & {#4} \end{array} \right)}
\newcommand\threemat[9]{\left(\begin{array}{ccc}  
{#1} & {#2} & {#3}\\ {#4} & {#5} & {#6}\\
{#7} & {#8} & {#9} \end{array} \right)}


\newcommand\cA{{\mathcal A}}
\newcommand\cB{{\mathcal B}}
\newcommand\cC{{\mathcal C}}
\newcommand\cD{{\mathcal D}}
\newcommand\cE{{\mathcal E}}
\newcommand\cF{{\mathcal F}}
\newcommand\cG{{\mathcal G}}
\newcommand\cH{{\mathcal H}}
\newcommand\cI{{\mathcal I}}
\newcommand\cJ{{\mathcal J}}
\newcommand\cK{{\mathcal K}}
\newcommand\cL{{\mathcal L}}
\newcommand\cM{{\mathcal M}}
\newcommand\cN{{\mathcal N}}
\newcommand\cO{{\mathcal O}}
\newcommand\cP{{\mathcal P}}
\newcommand\cQ{{\mathcal Q}}
\newcommand\cR{{\mathcal R}}
\newcommand\cS{{\mathcal S}}
\newcommand\cT{{\mathcal T}}
\newcommand\cU{{\mathcal U}}
\newcommand\cV{{\mathcal V}}
\newcommand\cX{{\mathcal X}}
\newcommand\cW{{\mathcal W}}
\newcommand\cY{{\mathcal Y}}
\newcommand\cZ{{\mathcal Z}}

\newcommand{\nit}{\noindent}
\newcommand{\ct}[1]{\cite{#1}}
\newcommand{\bib}[1]{\bibitem{#1}}

\newcommand\PRL[3]{\textsl{Phys. Rev. Lett.} \textbf{#1} (#2) #3}
\newcommand\NPB[3]{\textsl{Nucl. Phys.} \textbf{B#1} (#2) #3}
\newcommand\NPBFS[4]{\textsl{Nucl. Phys.} \textbf{B#2} [FS#1] (#3) #4}
\newcommand\CMP[3]{\textsl{Commun. Math. Phys.} \textbf{#1} (#2) #3}
\newcommand\PRD[3]{\textsl{Phys. Rev.} \textbf{D#1} (#2) #3}
\newcommand\PLA[3]{\textsl{Phys. Lett.} \textbf{#1A} (#2) #3}
\newcommand\PLB[3]{\textsl{Phys. Lett.} \textbf{#1B} (#2) #3}
\newcommand\CQG[3]{\textsl{Class. Quantum Grav.} \textbf{#1} (#2) #3}
\newcommand\JMP[3]{\textsl{J. Math. Phys.} \textbf{#1} (#2) #3}
\newcommand\PTP[3]{\textsl{Prog. Theor. Phys.} \textbf{#1} (#2) #3}
\newcommand\SPTP[3]{\textsl{Suppl. Prog. Theor. Phys.} \textbf{#1} (#2) #3}
\newcommand\AoP[3]{\textsl{Ann. of Phys.} \textbf{#1} (#2) #3}
\newcommand\RMP[3]{\textsl{Rev. Mod. Phys.} \textbf{#1} (#2) #3}
\newcommand\PR[3]{\textsl{Phys. Reports} \textbf{#1} (#2) #3}
\newcommand\FAP[3]{\textsl{Funkt. Anal. Prilozheniya} \textbf{#1} (#2) #3}
\newcommand\FAaIA[3]{\textsl{Funct. Anal. Appl.} \textbf{#1} (#2) #3}
\newcommand\TAMS[3]{\textsl{Trans. Am. Math. Soc.} \textbf{#1} (#2) #3}
\newcommand\InvM[3]{\textsl{Invent. Math.} \textbf{#1} (#2) #3}
\newcommand\AdM[3]{\textsl{Advances in Math.} \textbf{#1} (#2) #3}
\newcommand\PNAS[3]{\textsl{Proc. Natl. Acad. Sci. USA} \textbf{#1} (#2) #3}
\newcommand\LMP[3]{\textsl{Letters in Math. Phys.} \textbf{#1} (#2) #3}
\newcommand\IJMPA[3]{\textsl{Int. J. Mod. Phys.} \textbf{A#1} (#2) #3}
\newcommand\IJMPD[3]{\textsl{Int. J. Mod. Phys.} \textbf{D#1} (#2) #3}
\newcommand\TMP[3]{\textsl{Theor. Math. Phys.} \textbf{#1} (#2) #3}
\newcommand\JPA[3]{\textsl{J. Physics} \textbf{A#1} (#2) #3}
\newcommand\JSM[3]{\textsl{J. Soviet Math.} \textbf{#1} (#2) #3}
\newcommand\MPLA[3]{\textsl{Mod. Phys. Lett.} \textbf{A#1} (#2) #3}
\newcommand\JETP[3]{\textsl{Sov. Phys. JETP} \textbf{#1} (#2) #3}
\newcommand\JETPL[3]{\textsl{ Sov. Phys. JETP Lett.} \textbf{#1} (#2) #3}
\newcommand\PHSA[3]{\textsl{Physica} \textbf{A#1} (#2) #3}
\newcommand\PHSD[3]{\textsl{Physica} \textbf{D#1} (#2) #3}
\newcommand\JPSJ[3]{\textsl{J. Phys. Soc. Jpn.} \textbf{#1} (#2) #3}
\newcommand\JGP[3]{\textsl{J. Geom. Phys.} \textbf{#1} (#2) #3}

\newcommand\Xdot{\stackrel{.}{X}}
\newcommand\xdot{\stackrel{.}{x}}
\newcommand\ydot{\stackrel{.}{y}}
\newcommand\yddot{\stackrel{..}{y}}
\newcommand\rdot{\stackrel{.}{r}}
\newcommand\rddot{\stackrel{..}{r}}
\newcommand\vpdot{\stackrel{.}{\varphi}}
\newcommand\vpddot{\stackrel{..}{\varphi}}
\newcommand\tdot{\stackrel{.}{t}}
\newcommand\zdot{\stackrel{.}{z}}
\newcommand\etadot{\stackrel{.}{\eta}}
\newcommand\udot{\stackrel{.}{u}}
\newcommand\vdot{\stackrel{.}{v}}
\newcommand\rhodot{\stackrel{.}{\rho}}
\newcommand\xdotdot{\stackrel{..}{x}}
\newcommand\ydotdot{\stackrel{..}{y}}



\title{Dynamical Couplings and Charge Confinement/Deconfinement from Gravity 
Coupled to Nonlinear Gauge Fields \thanks{Talk at Second Bulgarian National 
Congress in Physics, Sept. 2013}}

\begin{start}
\author{E.~Guendelman}{1}, \coauthor{A.~Kaganovich}{1},
\coauthor{E.~Nissimov}{2}, \coauthor{S.~Pacheva}{2}

\address{Department of Physics, Ben-Gurion Univ. of the Negev,
Beer-Sheva 84105, Israel}{1}

\address{Institute of Nuclear Research and Nuclear Energy,
Bulg. Acad. Sci., Sofia 1784, Bulgaria}{2}

\runningheads{E.~Guendelman, A.~Kaganovich, E.~Nissimov, S.~Pacheva}{Dynamical 
Couplings and Charge Confinement/Deconfinement $\ldots$}

\received{}

\begin{Abstract}
We briefly outline several main results concerning various new physically relevant
features found in gravity -- both ordinary Einstein or $f(R)=R+R^2$ 
gravity in the first-order formalism, coupled to a special kind of nonlinear 
electrodynamics containing a square-root of the standard Maxwell Lagrangian and known 
to produce charge confinement in flat spacetime.
\end{Abstract}

\PACS{04.50.-h,04.70.Bw,11.25.-w}

\end{start}

\section[]{Introduction}

G. 't Hooft \ct{thooft} has shown that in any effective quantum gauge theory, which is 
able to describe QCD-like charge confinement phenomena, the energy density of 
electrostatic field configurations should be a linear function of the electric 
displacement field in the infrared region (the latter appearing as a quantum
``infrared counterterm''). The simplest way to realize these ideas in flat spacetime
is to incorporate into the full gauge field action an additional term being a 
square-root of the standard Maxwell (or Yang-Mills) gauge field Lagrangian 
\ct{GG-1,GG-2,GG-3}:
\br
S = \int d^4 x L(F^2) \quad ,\quad
L(F^2) = -\frac{1}{4} F^2 - \frac{f_0}{2} \sqrt{-F^2} \; ,
\lab{GG-flat} \\
F^2 \equiv F_{\m\n} F^{\m\n} \quad ,\quad 
F_{\m\n} = \pa_\m A_\n - \pa_\n A_\m  \; .
\nonu
\er
The ``square-root'' Maxwell term is naturally produced as a result of 
spontaneous breakdown of scale symmetry in the standard gauge theory \ct{GG-1}.
Moreover, the (dimensionfull) coupling constant $f_0$ measures the strength
of the effective confining potential among quantized fermions produced by
\rf{GG-flat} \ct{GG-2}.

In a series of recent papers \ct{our-main-1,our-main-2} we have studied in detail
physically more 
interesting models where the above nonlinear ``square-root electrodynamics'' 
couples to gravity --– either standard Einstein gravity or generalized 
$f(R)$-gravity ($f(R)$ being a nonlinear function of the scalar curvature of 
space-time), as well as coupled to scalar dilaton field. Let us recall that 
$f(R)$-gravity models are attracting a lot of interest in modern cosmology as 
possible candidates to cure problems in the standard cosmological scenarios 
related to dark matter and dark energy. For a recent review, see 
\textsl{e.g.} \ct{f(R)-grav} and references therein. The first $R+R^2$-model 
(in the second-order formalism) which was also the first inflationary model, 
was proposed by Starobinsky in \ct{starobinsky}.

Here we will briefly describe some of our main results \ct{our-main-1,our-main-2} 
concerning the new physically relevant features we uncovered in the coupled 
gravity/``square-root'' nonlinear gauge field/dilaton system (defined in 
Eq.\rf{f-gravity+GG+D} below):

(i) Appearance of dynamical effective gauge couplings and confinement-deconfinement 
transition effect as functions of the dilaton vacuum expectation value, in
particular due to appearance of ``flat'' region of the effective dilaton
potential.

(ii) New mechanism for dynamical generation of cosmological constant.

(iii) Non-standard black hole solutions with constant vacuum radial electric field 
with Reissner-Nordstr{\"o}m-(anti)de-Sitter or Schwarzschild-(anti)de-Sitter type 
geometry and with non-asymptotically flat ``hedgehog''-type spacetime asymptotics.
Let us stress that constant vacuum radial electric fields do not exist as
solutions of ordinary Maxwell electrodynamics.

(iv) The above non-standard black holes obey the first law of black hole 
thermodynamics.

(v) New ``tube-like universe''" solutions of Levi-Civita-Bertotti-Robinson tye
\ct{LC-BR};

(vi) Coupling to {\em lightlike} branes produces {\em "charge-hiding}" and 
{\em charge-confining} ``thin-shell'' wormhole solutions displaying QCD-like 
charge confinement (see also the previous talk \ct{BJP-LL-branes} at this conference).

\section{$R+R^2$-Gravity Coupled to Confining Nonlinear Gauge Field}

Let us consider coupling of $f(R)= R + \a R^2$ gravity (possibly
with a bare cosmological constant $\L_0$) to a ``dilaton'' $\phi$ and
the nonlinear gauge field system containing $\sqrt{-F^2}$ \rf{GG-flat} (we
are using units with the Newton constant $G_N=1$):
\br
S = \int d^4 x \sqrt{-g} \Bigl\lb \frac{1}{16\pi} 
\Bigl( f\bigl(R(g,\G)\bigr) - 2\L_0 \Bigr) + L(F^2(g)) + L_D (\phi,g) \Bigr\rb \; ,
\lab{f-gravity+GG+D} \\
f\bigl(R(g,\G)\bigr) = R(g,\G) + \a R^2(g,\G) \quad ,\quad 
R(g,\G) = R_{\m\n}(\G) g^{\m\n} \; ,
\lab{f-gravity} \\
L(F^2(g)) = - \frac{1}{4e^2} F^2(g) - \frac{f_0}{2} \sqrt{- F^2(g)} \; ,
\lab{GG-g} \\
F^2(g) \equiv F_{\k\l} F_{\m\n} g^{\k\m} g^{\l\n} \;\; ,\;\;
F_{\m\n} = \pa_\m A_\n - \pa_\n A_\m \;
\lab{F2-g} \\
L_D (\phi,g) = -\h g^{\m\n}\pa_\m \phi \pa_\n \phi - V(\phi) \; .
\lab{L-dilaton}
\er
$R_{\m\n}(\G)$ is the Ricci curvature in the first order (Palatini) formalism, 
\textsl{i.e.}, the space-time metric $g_{\m\n}$ and the affine connection 
$\G^\m_{\n\l}$ are \textsl{a priori} independent variables. The solution to the
corresponding equation of motion w.r.t. $\G^\m_{\n\l}$ --
$\nabla_\l \(\sqrt{-g} f^{\pr}_R g^{\m\n}\) = 0$ -- implements transition 
to the ``physical'' Einstein-frame metrics $h_{\m\n}$ via conformal rescaling of 
the original metric $g_{\m\n}$:
\be
g_{\m\n} = \frac{1}{f^{\pr}_R} h_{\m\n} \quad ,\quad
\G^\m_{\n\l} = \h h^{\m\k} \(\pa_\n h_{\l\k} + \pa_\l h_{\n\k} - \pa_\k h_{\n\l}\)
\; .
\lab{einstein-frame}
\ee
Here $f^{\pr}_R \equiv \frac{df(R)}{dR} = 1 + 2\a R(g,\G)$.

As shown in \ct{our-main-2}, using \rf{einstein-frame} the original $R+R^2$-gravity 
equations of motion resulting from \rf{f-gravity+GG+D} can be
rewritten in the form of {\em standard} Einstein equations:
\be
R^\m_\n (h) = 8\pi \({T_{\rm eff}}^\m_\n (h) - \h \d^\m_\n {T_{\rm eff}}^\l_\l (h)\)
\lab{einstein-h-eqs}
\ee
with effective energy-momentum tensor of the following form:
\be
{T_{\rm eff}}_{\m\n} (h) = h_{\m\n} L_{\rm eff} (h) 
- 2 \partder{L_{\rm eff}}{h^{\m\n}} \; .
\lab{T-h-eff}
\ee
The pertinent effective ``Einstein-frame'' matter Lagrangian reads:
\br
L_{\rm eff} (h) = - \frac{1}{4 e_{\rm eff}^2 (\phi)} F^2(h) 
- \h f_{\rm eff} (\phi) \sqrt{- F^2(h)} 
\nonu \\
+ \frac{X(\phi,h)\bigl(1+16\pi\a X(\phi,h)\bigr) - V(\phi) -\L_0/8\pi
}{1+8\a\( 8\pi V(\phi)+\L_0\)}
\lab{L-eff-h}
\er
with the following dynamical $\phi$-dependent couplings:
\br
\frac{1}{e_{\rm eff}^2 (\phi)} = \frac{1}{e^2} + 
\frac{16\pi\a f_0^2}{1 + 8\a \(8\pi V(\phi) + \L_0 \)} \; ,
\lab{e-eff} \\
f_{\rm eff}(\phi)=f_0 \frac{1+32\pi\a X(\phi,h)}{1 + 8\a\(8\pi V(\phi)+\L_0\)}
\; .
\lab{f-eff}
\er
The dilaton kinetic term $X(\phi,h) \equiv -\h h^{\m\n}\pa_\m \phi \pa_n \phi$ 
will be ignored in the sequel.

\section{Dynamical Couplings and Confinement/Deconfinement in $R+R^2$ 
Gravity}

In what follows we consider constant ``dilaton'' $\phi$ extremizing the effective
Lagrangian \rf{L-eff-h}. Here we observe an interesting feature of \rf{L-eff-h} --
the dynamical couplings and effective potential are extremized 
{\em simultaneously}, which is an explicit realization of the so called 
``least coupling principle'' of Damour-Polyakov \ct{damour-polyakov}:
\be
\partder{f_{\rm eff}}{\phi} = - 64\pi\a f_0 \partder{V_{\rm eff}}{\phi}
\;\; ,\;\; \partder{}{\phi}\frac{1}{e_{\rm eff}^2} =
-(32\pi\a f_0)^2 \partder{V_{\rm eff}}{\phi} 
\;\; \to \partder{L_{\rm eff}}{\phi} \sim \partder{V_{\rm eff}}{\phi} \; ,
\lab{f-e-extremize}
\ee
where:
\be
V_{\rm eff}(\phi) = 
\frac{V(\phi) + \frac{\L_0}{8\pi}}{1+8\a\(8\pi V(\phi)+\L_0\)} \; .
\lab{V-eff}
\ee
Thus, at a constant extremum of $L_{\rm eff}$ \rf{L-eff-h} $\phi$ must satisfy:
\be
\partder{V_{\rm eff}}{\phi} = 
\frac{V^{\pr}(\phi)}{\llb 1+8\a\( 8\pi V(\phi)+\L_0\)\rrb^2} = 0 \; .
\lab{V-extremum}
\ee
There are two generic cases:

(a) {\em Confining phase}: Eq.\rf{V-extremum} is satisfied for some 
finite-value $\phi_0$ extremizing the original ``bare'' dilaton potential $V(\phi)$: 
$V^{\pr}(\phi_0) = 0$.

(b) {\em Deconfinement phase}: For polynomial or exponentially growing ``bare'' 
potential $V(\phi)$, so that $V(\phi) \to \infty$ when $\phi \to \infty$, we have:
\be
\partder{V_{\rm eff}}{\phi} \to 0 \quad ,\quad 
V_{\rm eff} (\phi) \to \frac{1}{64\pi\a} = {\rm const} \quad {\rm when} \;\;
\phi \to \infty \; ,
\lab{flat-region}
\ee
\textsl{i.e.}, for sufficiently large values of $\phi$ we find a ``flat region''
in $V_{\rm eff}$. This ``flat region'' triggers a {\em transition from 
confining to deconfinement dynamics}.

Namely, in the ``flat-region'' phase (b) we have 
$f_{\rm eff} \to 0 \; ,\; e^2_{\rm eff} \to e^2$,
and the effective gauge field Lagrangian in \rf{L-eff-h} reduces to the ordinary
\textsl{non-confining} one (the ``square-root'' Maxwell term $\sqrt{-F^2}$ vanishes):
\be
L^{(0)}_{\rm eff} = -\frac{1}{4e^2} F^2(h) - \frac{1}{64\pi\a}
\lab{L-eff-h-0}
\ee
with an {\em induced} cosmological constant $\L_{\rm eff} = 1/8\a$, which is
{\em completely independent} of the bare cosmological constant $\L_0$.

\section{Non-Standard Black Holes and ``Tube-Like'' Universes}

In the {\em confining phase} (phase (a) above: 
$\phi_0$ -- generic minimum of the effective dilaton potential; 
$e_{\rm eff}(\phi))$, $f_{\rm eff}(\phi)$ as in \rf{e-eff}--\rf{f-eff})
we obtain several physically interesting 
solutions of the Einstein-frame gravity Eqs.\rf{einstein-h-eqs} due to the special 
form of the effective Einstein-frame matter Lagrangian \rf{L-eff-h}.

First, we find \textsl{non-standard} Reissner-Nordstr{\"o}m-(anti-)de-Sitter-type
black holes with metric:
\br
ds^2 = - A(r) dt^2 + \frac{dr^2}{A(r)} + r^2 \bigl(d\th^2 + \sin^2 \th d\vp^2\bigr)
\; ,
\lab{spherical-static-2} \\
A(r) = 1 - \sqrt{8\pi}|Q|f_{\rm eff}(\phi_0) e_{\rm eff}(\phi_0) 
- \frac{2m}{r} + \frac{Q^2}{r^2} - \frac{\L_{\mathrm{eff}}(\phi_0)}{3} r^2 \; ,
\lab{CC-eff-2}
\er
where $\L_{\rm eff}(\phi_0)$ is a {\em dynamically generated} cosmological constant:
\be
\L_{\rm eff}(\phi_0) = \frac{\L_0 +8\pi V(\phi_0)+2\pi e^2 f^2_0}{
1+8\a\(\L_0 +8\pi V(\phi_0)+2\pi e^2 f^2_0\)} \; .
\lab{h-CC-eff}
\ee
The black hole's static radial electric field contains
apart from the Coulomb term an {\em additional constant ``vacuum'' piece}:
\br
|F_{0r}| = |\vec{E}_{\rm vac}| -
\frac{Q}{\sqrt{4\pi}\, r^2} \Bigl(\frac{1}{e^2} + 
\frac{16\pi\a f_0^2}{1 + 8\a\(8\pi V(\phi_0) + \L_0 \)}\Bigr)^{-\h}
\lab{cornell-sol-2} \\
|\vec{E}_{\rm vac}| \equiv 
\Bigl(\frac{1}{e^2} + 
\frac{16\pi\a f_0^2}{1 + 8\a\(8\pi V(\phi_0) + \L_0 \)}\Bigr)^{-1}
\frac{f_0/\sqrt{2}}{1+8\a\(8\pi V(\phi_0)+\L_0\)} \; .
\lab{vacuum-radial}
\er
Let us emphasize again that constant vacuum radial electric fields do not exist as
solutions of ordinary Maxwell electrodynamics.

In the special case $\L_{\rm eff}(\phi_0)=0$ we obtain a non-standard
Reissner-Nordstr{\"o}m-type black hole with a ``hedgehog'' 
{\em non-flat-spacetime} asymptotics (cf. Refs.\ct{hedgehog}:
$A(r) \to 1 -\sqrt{8\pi}|Q|f_{\rm eff}(\phi_0) e_{\rm eff}(\phi_0) \neq 1$
for $ r\to\infty$.

Apart from non-standard black hole solutions we also obtain ``tube-like''
space-time solutions of the Einstein-frame gravity Eqs.\rf{einstein-h-eqs},
which are of Levi-Civitta-Bertotti-Robinson type (cf. Refs.\ct{LC-BR}) with 
geometries $AdS_2 \times S^2$, $Rind_2 \times S^2$ and $dS_2 \times S^2$,
where $AdS_2$, $Rind_2$ and $dS_2$ are two-dimensional anti-de Sitter,
Rindler and de Sitter space, respectively. The corresponding metric is of the form:
\be
ds^2 = - A(\eta) dt^2 + \frac{d\eta^2}{A(\eta)} 
+ r_0^2 \bigl(d\th^2 + \sin^2 \th d\vp^2\bigr) \;\; ,\;\;  
-\infty < \eta <\infty \; ,
\lab{gen-BR-metric-2}
\ee
carrying constant vacuum ``radial'' electric field $|F_{0\eta}|=|\vec{E}_{\rm vac}|$.
The radius of the spherical factor $S^2$ is given by 
(using short-hand $\L(\phi_0)\equiv 8\pi V(\phi_0) + \L_0$):
\be
\frac{1}{r_0^2} = \frac{4\pi}{1+8\a\L(\phi_0)}\Bigl\lb
\Bigl(1+8\a\(\L(\phi_0)+2\pi f_0^2\)\Bigr) \vec{E}_{\rm vac}^2 +
\frac{1}{4\pi}\L(\phi_0)\Bigr\rb \; ,
\lab{r0-eq}
\ee
and the metric coefficient in \rf{gen-BR-metric-2} reads:
$A(\eta) = 4\pi K(\vec{E}_{\rm vac}) \eta^2 \;,\; K(\vec{E}_{\rm vac}) >0$ for $AdS_2$;
$A(\eta) = \pm \eta$ for $\eta \in (0,+\infty)$ or  $\eta \in (-\infty,0)$
for $Rind_2$;
$A(\eta) = 1 - 4\pi |K(\vec{E}_{\rm vac})|\,\eta^2\; ,\;  K(\vec{E}_{\rm vac}) <0$ 
for $dS_2$, using notation: \\
$K(\vec{E}) \equiv \Bigl(1+8\a\(\L(\phi_0)+2\pi f_0^2\)\Bigr) \vec{E}^2 
- \sqrt{2}f_0 |\vec{E}| - \frac{1}{4\pi}\L(\phi_0)$.

To conclude, let us particularly stress that all physically relevant
features described above are exclusively due to the {\em combined} effect of
both square-root nonlinearity $\sqrt{-F^2}$ in the gauge field Lagrangian as well 
as the $R^2$ term in the gravity action.



\section*{Acknowledgments}
We gratefully acknowledge support of our collaboration through the academic exchange 
agreement between the Ben-Gurion University and the Bulgarian Academy of Sciences.
S.P. has received partial support from COST action MP-1210.


\end{document}